\documentclass[%
 aip,
 jmp,%
 amsmath,amssymb,
 reprint,%
]{revtex4-1}

\usepackage{graphicx}
\usepackage{dcolumn}
\usepackage{bm}
\usepackage{color}
\usepackage{hyperref}
\hypersetup{colorlinks=true,linkcolor=red,citecolor=red,urlcolor=red}
\usepackage{upgreek}
\usepackage{epstopdf} 

\begin{document}

\preprint{-}

\title[{S.R Mirfayzi et al. }]{A Non-linear Reaction-Diffusion System in Propagating Diffusive Wave in Nitroguanidine (NQ) Lens Using Ising-Bloch Bifurcation}

\author{S.R. Mirfayzi}
 \altaffiliation{Department of Maths \& Physics, Queen's University Belfast, Belfast, BT71NN, UK.}
 \email{smirfayzi01@qub.ac.uk}


\begin{abstract}
A wave front propagating through a medium is described using the Ising--Bloch method The reaction-diffusion behaviour of an autocatalysis model of incoming waves from energetic material in a nitroguanidine lens and its interactions with Dirichlet boundaries system are examined. Wave splitting is found to occur for some relative diffusivities through introducing defects into the lens. The system is introduced using a nonlinear method with a Boltzmann--Gibbs distribution of diffusion parameters and forced into a right-hand-side divergence for final analysis. Considering the nonlinearity, the model is expanded by introducing the decomposition kinetics into the set of equations developed.

\end{abstract}

\maketitle

\begin{quotation}
Statement of Purpose:\newline
The dynamics of waves propagating in medium can be constructed and designed to exhibit noble statement of purpose using fixed chemical points. The real-world models should hierarchically organized into layers of modules and sub-modules of these chemicals. Examples are the creation of active/passive lenses and shields, where waves are guided with an externally applied field through interfaced chemicals within the medium, so that the diffusive products, compression waves and stress waves can be guided and pre-determined. The idea developed in this paper is unique and has never had done before and it shows the properties of such a system within a non-linear reaction diffusion model.    
\end{quotation}

\section{\label{sec:level1}Introduction \lowercase{} }

\begin{figure}[ht!]
\centering
\includegraphics[width=90mm]{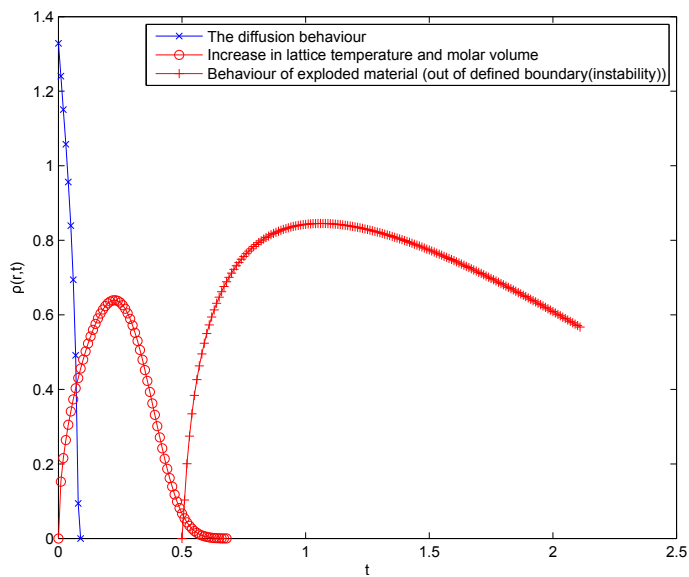}
\caption{Diffusion in nitroguanidine for averaged $N_2$ and C$-$N molecules. At $t=0.01$ to $t=0.08$, there is an increase temperature due to incoming waves; At $t=0.08$ to $t=0.5$, there is a phase change from solid to vapor, where at $t=0.5$ outburst occurs ($t=0.5$ is defined outside of the Dirichlet boundaries.)}
\label{arbode}
\end{figure}

The reaction diffusion system studied in this paper is described using an established analytical approach,\cite{hagberg1997dynamics, barkley1991model, mikhailov1990foundations, PhysRevLett.60.1880, PhysRevE.51.1899, hagberg1994complex}. The nonuniform motion of the solution front is also investigated. The solution of front is given by a relationship between the normal velocity of the front vector and its curvature $k$ consisting of an integrodifferential equation for the front curve.\cite{hagberg1997dynamics} The model is a simple, two-component model that gives rise to a fast propagator (activator) and recovery variable known as an inhibitor.\cite{winfree2001geometry, murray2002mathematical, field1985oscillations} This is due to presence of nonequilibrium Ising-Bloch bifurcation. \newline 
Adding the kinetic coefficients requires detailed reaction schemes of incoming waves from the energetic material into the lens. The basic mechanisms are: phase change to solid intermediates and the formation of gaseous products. Because the material decomposes below its melting point, this results in a complex pathway involving NO, nitroso, and Non-Violent Resistance (NVR) cycles. One pathway involves elimination of HONO and HON, leading to the formation of HCN  (OST; C$_3$H 3N$_3$O) under high confinement conditions. The other cycle includes the formation of H$_2$O, NO and NO$_2$. The next reaction leads to NO cycles, NO$_2$ with nitroso cycles and the formation of minor products at higher pressures,\cite{behrens1991thermali} Interfacing the material with nitroguanidine leads to diffusion of decomposition products and the creation of spiral waves in the chemical. Figure \ref{arbode} demonstrates the  behaviour of nitroguanidine on a model developed for this purpose based on a nonlinear system. 
The stars show the diffusion of $N_2$ molecules into nitroguanidine, which causes a temperature increase creating an unstable condition within the system. This leads to an increase in kinetic energy and molar volume, and ultimately to the explosion of the chemical. This work will lead to important new research in the fields of guided stress/shock waves.
\section{\label{sec:level1}The Model \lowercase{} }
The model developed can be applied to any diffusive products in organic reactant. In this paper, the diffusive products are dealt with as external control parameters which affect the wave behaviour in the lens. Specifically, these parameters are the crystal temperature and partial pressures of the decomposition products. Hence, an understanding of the decomposition is essential. In studying the decomposition of such a system, it is always assumed that it enters the vapor phase before dissociating. In a series of experiments, It was found to release about 500 kJ.mol$^{-1}$ of energy, independent of the heating rate and mass. The majority of the decomposition products form via the scission of $N_2$, N-NO$_2$ and C$-$N bonds. The decomposition is largely independent of the heating rate, which is due to the total heat release (due to the exothermic decomposition reactions). Finding bifurcation requires knowledge of the partial pressures of the reactants and products, as well as their kinetic rates.\cite{park1993kinetics}. According to several study the kinetic behaviour of molecular dissociation for $N_2$ bond dissociation is 200 kJ/mol and the energy of decomposition of the triple bond-fission reaction products of CH$_2$NNO$_2$ are 155, 234 and 297 kJ.mol$^{-1}$. 
\newline
For the lens, the reaction-diffusion system is introduced through the bond-fission products of $N_2$  and C$-$N at Dirichlet boundaries. Because this is a bistable system, a front-like structure connects the two fronts in a homogeneous steady state. A small perturbation known as an Ising--Bloch bifurcation\cite{yadav2004interaction} exists, where the fronts exchange stability with a pair of counter-propagating Bloch fronts. This front motion is broken by imposing a fixed chemical concentration at the boundary of the reactor. Here, the nitroguanidine molecule is considered as a stationary spot to bifurcate the incoming fronts. The spiral waves and breaking spots in an excitable\cite{meron1992pattern} and bistable medium are examples of oscillatory behaviour; the oscillations often result from the underlying oscillating dynamics of local chemical kinetics.\cite{haim1996breathing}\newline
\begin{figure*}
\centering
\begin{minipage}[ht!]{.5\textwidth}\centering
\includegraphics[scale=0.50]{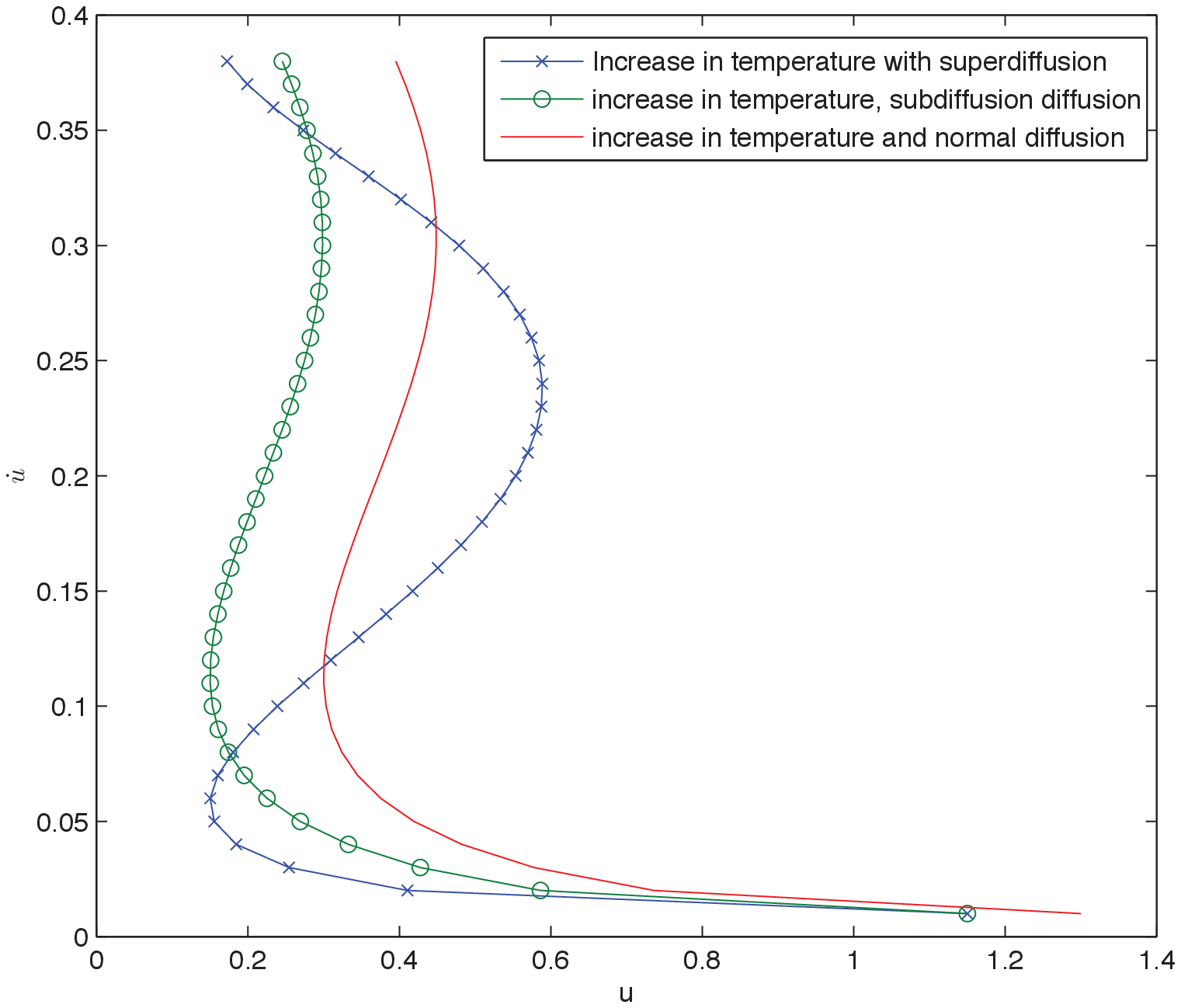}
\end{minipage}\hfill
\begin{minipage}[ht!]{.5\textwidth}\centering
\includegraphics[scale=.90]{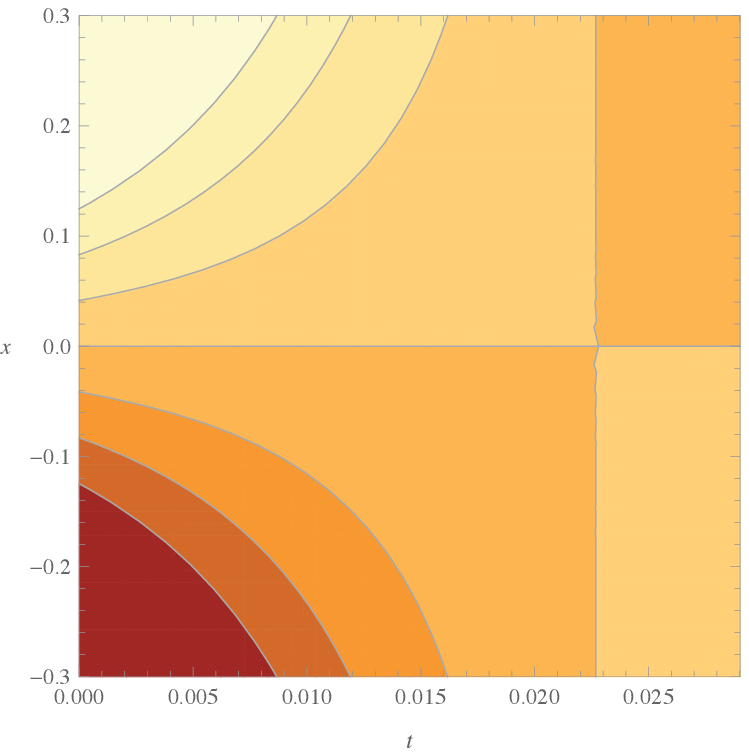}
\end{minipage}\hfill
\caption{a) Solution to Eqs. 1 and 2 for different values of the diffusion parameter $D$. The data show the increase in temperature in the lattice. b) Compression front at the nitroguanidine interface boundary.}
\label{difphase1}
\end{figure*}
The model presented in this paper was used to study the wave propagation under three dynamic boundary conditions. This propagation can influence the state of reactants by displacing them. This is due to the increase in specific volume, which generates compression waves and sets the reactants and flame front into motion following precursor shock. Thus, the three important, overriding factors are the wave speed, specific volume ratio and boundary conditions. The boundary condition in this model is therefore set at a critical speed of $(S/\alpha)$ at which the initial wave speed is lower than that of the shockwave-induced compression ($M_s$), such that an acceleration mechanism can be introduced through an external field. Here, $S$ is the burn velocity, and $\alpha$ is sound speed in front of the shock. \newline
Under this condition, the chemical reaction is affected by the compression shock wave and shock diffusion replaces diffusion-controlled flame propagation.\cite{spitzer2010energetic,krischer1992oscillatory,hagberg1994pattern}      
Having established the following parameters: crystal pressure (T$_c$), partial pressures for $N_2$ and C$-$N resulting from bond dissociation as P$_{N_2}$ and P$_{CN}$, kinetic rates ($k$), surface area of the reaction diffusion system (S$_a$) and the external field ($h$), the reaction diffusion system is given by:  
\begin{equation}	
\dot{u}=\epsilon^{-1}(-k_d u- a_0 u^3-k_r u)+ \delta_t \rho_{xx}
\label{bif1}
\end{equation}
and
\begin{equation}	
\dot{v}=u- a_1( k_r v - \delta v_{xx} ) + h
\label{bif2}
\end{equation}
where $u$ and $v$ are the two variables for the propagator and inhibitor.\cite{bar2009spiral,bode1994measurement,haas1995observation} The parameter $a_0$ represents the kinetic expression for $k_u. S_a. P_{N_2}$ and $a_1$ represents $k_u. S_{v}. P_{CN}$, where $k_u$ is given by the rate of molecules hitting the surface for P$_{N_2}$ and P$_{CN}$ (note that $u$ is given in units of saturation coverage), $P_{N_2}$ and $P_{CN}$ are the partial pressures induced by bond dissociation and $S_a$ and $S_{v}$ are the lattice cross-sections. $k_d$ and $k_r$ represent the desorption and reaction rates. The reaction rate and other parameters were taken from mechanistic studies based on a comparison between experimental data and theoretical modeling using the Arrhenius equation.\cite{huang1987detection, maharrey2005thermal,melius1987j}
In Eq. 2, $h$ represents the induced compression wave. The system has two solutions for the up and down states. The localized functions $a_0$ and $a_1$ are defined through the decomposition properties of the material. However, by arbitrary definition, if $a_0=0$, the stationary front solution loses its stability to Pitchfork bifurcation for $\epsilon > \epsilon_c( \delta)$.\cite{hagberg1996controlling} \newline
Arbitrary analysis of Eq. \ref{bif1} and \ref{bif2} reveals a bistability.
The parameter $\delta$ affects the relative spatial extent of the front.
$D$ in Eq. 3 is composed of the diffusion parameter of diffusive products in the nitroguanidine crystal and is given in a nonlinear form through a Boltzmann--Gibbs distribution:
\begin{equation}	
\delta_t \rho_{xx} =D\bigtriangledown^2 v_{xx}
\label{bif3}
\end{equation}
The anomalous diffusion equation is dependent on how it is applied. Here, the diffusion model was developed through a nonlinear system.\cite{plastino1995non} In this case because of the introduction of external forces to the reaction-diffusion system, the same approach has to be applied to the diffusion parameter. The nonlinear system defined is based on a Tsallis formalism or a standard thermostatic Boltzmann--Gibbs equation\cite{tsallis1996anomalous, huang1987statistical} for diffusion of the decomposition products in lenses. Therefore, $v_{xx}$ is defined by\cite{lenzi2005nonlinear}:
\begin{equation}	
v_{xx}(r)=r^{-2} exp_q[-\L(r)/z]
\label{bifdif1}
\end{equation}       
The term $exp_q$ exists for two conditions of $q=1$ and $q>1$ \cite{plastino1995non}. If right-hand-side divergence of the equation $v_{xx}(r)$ is assumed, the function at $q=1$ converges to the ordinary differential equation exponential function.\cite{naudts2010q}
The function $\L(r)$ is given in ref. \cite{assis2006nonlinear}. Assuming external forces are present, the function becomes:
\begin{equation}	
v_{xx}(r)=r^{-2} exp((\frac{1}{D}) \int_0^r d\hat{r}-kr)
\label{bifdif2}
\end{equation}   
With the exertion of external forces, the equation above needs to satisfy the developed boundary condition for the explicit time-dependent equation given by:
\begin{equation}	
v_{xx}(r,t)= (\frac{1}{\phi(t)})(\frac{r}{\phi(t)})^{\frac{D}{2} -1} exp[(-\frac{1}{0.65D^2})(\frac{r}{\phi(t)})^{1.3D}]
\label{bifdif3}
\end{equation}   
The solution to $\phi(t)$ is given in ref. \cite{assis2006nonlinear}. The diffusion parameter $D$ affects the total system behaviour, where $1<D<2$, $2<D<3$ or $D>3$, and leads to subdiffusive behaviour at the boundary and superdiffusive behaviour far from the edges. A simplified phase diagram for the relative diffusion parameter is given by figure \ref{difphase1}.\newline
The effect of external field can be defined for three regions: $(i)$ at the front corresponding to the down state invading the up state, $(ii)$ at the front corresponding to the up state invading the down state and $(iii)$ where both fronts exist. For a quantitative study of the front dynamics and relative pattern characteristics for the inner region, the front is defined at $x \rightarrow r= x-x_f(t)$, stretching the spatial coordinates according to $z=r/ \sqrt{2}$ and expanding $u$ and $v$ according to $\sqrt{\epsilon}$ steps. Here, $\epsilon$ corresponds to the ratio across the fronts. The initial value of $v$ was measured at the material interface with the nitroguanidine lens and $u$ was measured using $-tanh(z \sqrt{2})$. With a step size of $\sqrt{\epsilon}$ we have:
\begin{equation}	
\delta^2_z u_1+u-3u_0^2 u_1=v_1-x_fu_0z
\label{bif4}
\end{equation}     
The solution to Eq. \ref{bif4} yields the nonequilibrium Ising-Bloch bifurcation (NIB) velocity, which is calculated for the energetic diffusive molecules using:
\begin{equation}	
C_0=\frac{-3}{\eta \sqrt{2}}v_f
\label{bif5}
\end{equation}     
The total normal front velocity can be found using:
\begin{equation}	
C_n=C_0-k
\label{bif6}
\end{equation}
where $k$ is the curvature induced by compression waves during outburst. By calculating the front velocity ($C_n$), the effect of an external field at the front residue can also be analysed. As previously mentioned, the compression wave is treated as an external field which accelerates the flame velocity and diffusion in the lens and forms a Pitchfork bifurcation. The compression wave front in a heterogeneous solid is more complex than in a homogeneous solid because of the presence of voids, grains and flaws.\cite{tarver1993energy} A good approximation of the compression wave shape function in a homogeneous system can be given using a similar front to those developed for diffusion systems. Figure \ref{difphase1}b demonstrates the homogeneous compression fronts introduced to the system.     
The compression waves interact with the lower and upper boundaries at $h \approx 0$, which is the termination point. A dynamic equation for $v_f$ requires the branches to be scaled spatially. By assuming un-stretched coordinates: 
\begin{equation}	
v_t-C_nv_y=u \pm (v) - a_1( k_r (v + v_{yy})- \beta(y+ \frac{x_f \sqrt{\epsilon}}{\eta} )+h
\label{bif7}
\end{equation}     
Here, $\beta$ represents the compression wave front. By substituting Eq. \ref{bif6} into Eq. \ref{bif7}:
\begin{equation}	
v_t=u \pm (v) - a_1( k_r (v + v_{yy})- \beta(y+ \frac{x_f \sqrt{\epsilon}}{\eta} )+h +C_n
\label{bif8}
\end{equation}     
Initially $C_n$ is approximated at $v(0,t)$.
\begin{figure}
\centering
\begin{minipage}[ht]{.5\textwidth}\centering
\includegraphics[trim= 5mm 0 0 0 , scale=0.50]{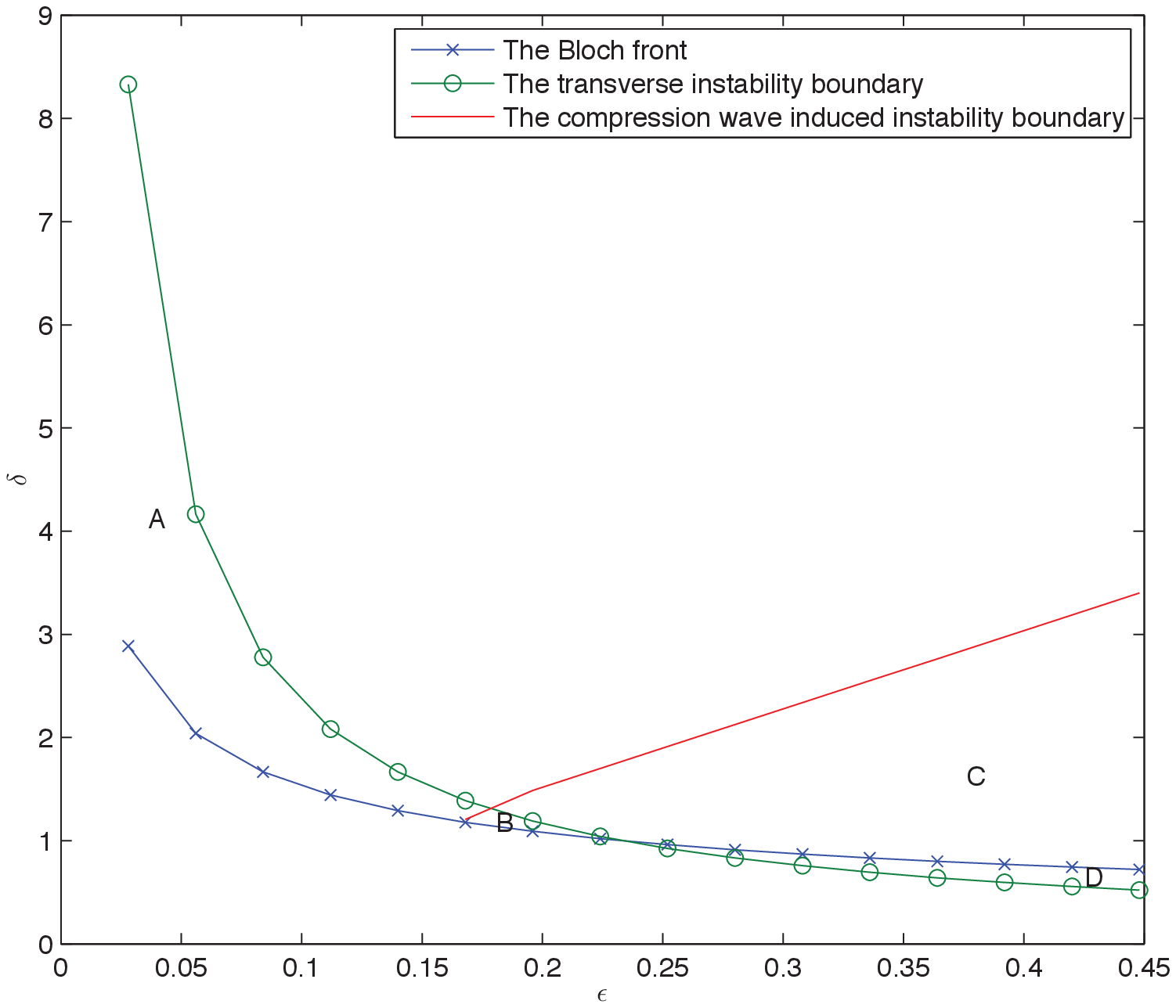}
\end{minipage}\hfill
\begin{minipage}[ht]{.5\textwidth}\centering
\includegraphics[trim= 5mm 0 0 0 , scale=0.48]{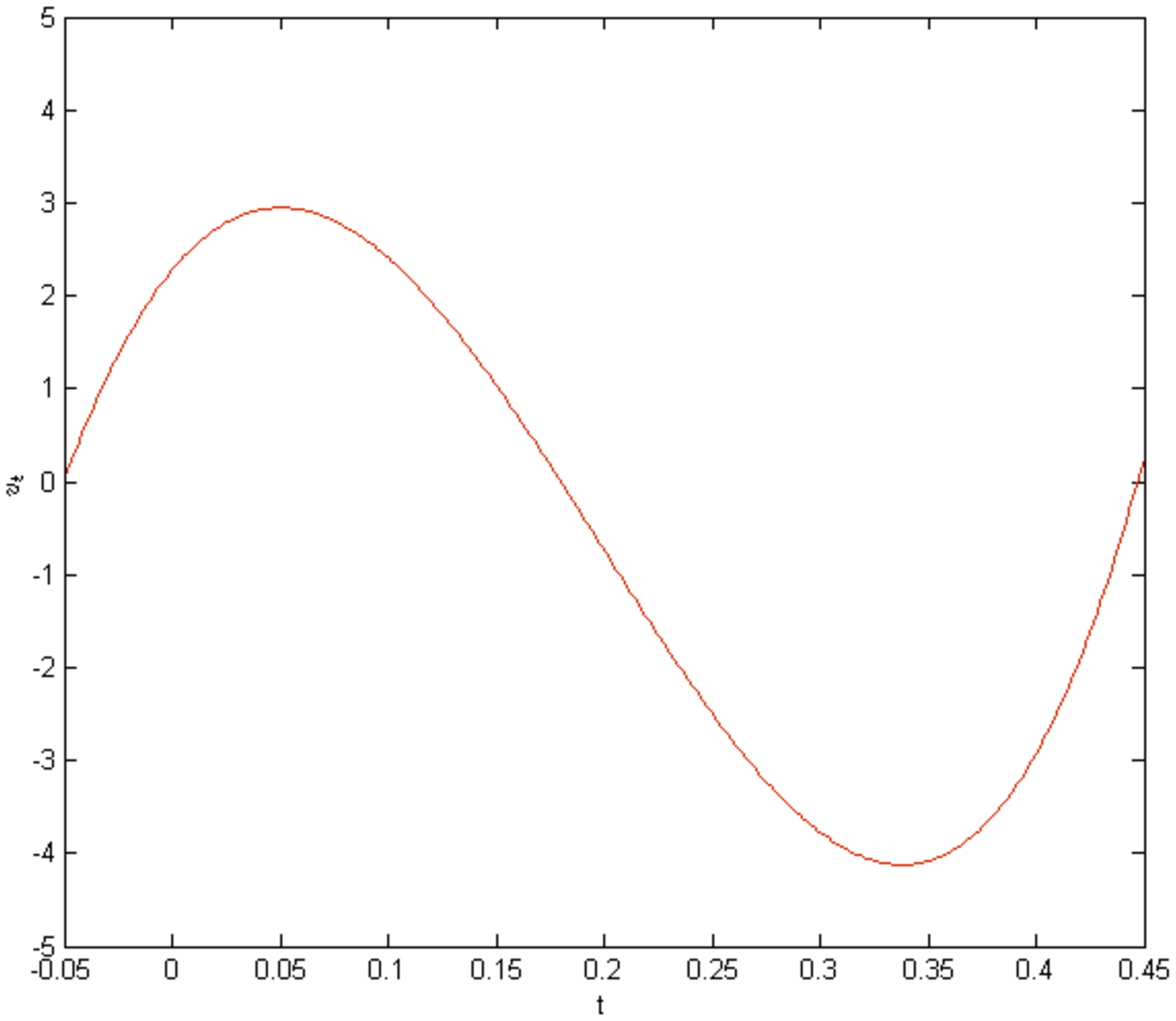}
\end{minipage}\hfill
\caption{a) Illustration of the front solution for Ising--Bloch (diamonds) and NIB bifurcation. The compression wave drives the system into instability through a metastable phase. b) Front solution as a function of time for an oscillatory unstable front.}
\label{difphase}
\end{figure}
The front solution to Eq. \ref{bif8}a is given for different diffusivity phases. Regions $(A)$ and $(D)$ show opposing behaviour of decreasing and increasing kinetics of the Bloch fronts through a metastable region $(B)$. Introducing the compression waves leads the system to instability (region $(D)$) causing faster energy release.  \newline  
Finally, oscillation in the model is produced so that the validity of the system for an unstable nonlinear condition can be approximated using Fig. \ref{difphase}b.

\section{\label{sec:level1}Discussion \lowercase{} }
The production of wave fronts in the nitroguanidine lens shows inhibitor production giving rise to two additional fronts at the chemical fixed point (nitroguanidine). This provides a predictive method for the future design of such a system where guided wave propagation in the medium is necessary. It was also shown that the front-like structure connects the two fronts in a homogeneous steady state along the fixed chemical point such that bifurcation occurs in the medium. However, the kinetics of phase transition in nitroguanidine are too complex to be described by a steady state form of equation because turbulent waves are formed upon the diffusion of decomposition at microseconds of diffusion, and fronts lose their stability to lattice deformation. However, an unstable solution could be obtained by introducing a sticking variable to Eq. \ref{bif2} to provide phase transformation instability to the solution, as demonstrated in Fig. \ref{difphase}a. Phase transformation therefore changes the faceting by influencing the adsorption rate, lattice heat increase, and extended induced partial bond dissociation. This also causes further external periodic perturbations in a metastable condition. \newline
Introducing the nonlinear form of diffusion increased the efficiency of such an approach if compared with the standard model.

\section{\label{sec:level1}Conclusion \lowercase{} }
The diffusion of the decomposition products in a nitroguanidine lens was studied in a chaotic system. The diffusion of the composites in a wave shaper was described by a two-variable model with inhibition and production in the reaction diffusion system. The calculation of wave propagation showed a good predictive behaviour for microseconds after the disruption in nitroguanidine.

\nocite{*}
\section{\label{sec:level1}References \lowercase{} }


\providecommand{\noopsort}[1]{}\providecommand{\singleletter}[1]{#1}%

\end{document}